# Brain Tumor Segmentation from MRI Images using Deep Learning Techniques

Ayan Gupta[1], Mayank Dixit[1], Vipul Kumar Mishra[2], Attulya Singh[1], Atul Dayal[1]

[1] Galgotias College of Engineering and Technology, Greater Noida, 201306, UP, India
[2] School of Computer Science Engineering & Technology, Bennett University, Greater Noida, India

```
{ayangupta.19gcebcs080;mayank.dixit}@galgotiacollege.edu,
vipul.mishra@bennett.edu.in,{attulyasingh19gcebcs121;
atuldatal.19gcebcs107}@galgotiacollege.edu
```

**Abstract.** A brain tumor, whether benign or malignant, can potentially be life threatening and requires painstaking efforts in order to identify the type, origin and location, let alone cure one. Manual segmentation by medical specialists can be time-consuming, which calls out for the involvement of technology to hasten the process with high accuracy. For the purpose of medical image segmentation, we inspected and identified the capable deep learning model, which shows consistent results in the dataset used for brain tumor segmentation. In this study, a public MRI imaging dataset contains 3064 TI-weighted images from 233 patients with three variants of brain tumor, viz. meningioma, glioma, and pituitary tumor. The dataset files were converted and preprocessed before indulging into the methodology which employs implementation and training of some well-known image segmentation deep learning models like U-Net & Attention U-Net with various backbones, Deep Residual U-Net, ResUnet++ and Recurrent Residual U-Net. with varying parameters, acquired from our review of the literature related to human brain tumor classification and segmentation. The experimental findings showed that among all the applied approaches, the recurrent residual U-Net which uses Adam optimizer reaches a Mean Intersection Over Union of 0.8665 and outperforms other compared state-of-the-art deep learning models. The visual findings also show the remarkable results of the brain tumor segmentation from MRI scans and demonstrates how useful the algorithm will be for physicians to extract the brain cancers automatically from MRI scans and serve humanity.

# 1. Introduction

The term "brain tumor" refers to a mass or proliferation of abnormal cells in the brain. and the human skull being a closed space leaves no room for such growth to prevail. Hence, this abnormal growth can lead to unexpected developments for the worse. Brain tumors occur in several varieties. Both benign and malignant brain tumors can begin in the brain as primary brain tumors and can also spread to nearby tissues, or, in the form of secondary brain tumors, cancer that started in another part of the body can move to the brain [1]. Imaging tests are vital in identifying if the tumor is primary or secondary. Magnetic resonance imaging (MRI) of the brain is the initial step in the tumor's diagnosis. MRI can create precise scans of the body using magnetic fields, as compared to the results from X-rays, which is also helpful in determining the size of the tumor which makes it the ideal method to identify a brain tumor as it also produces images that are more precise than CT scans [2]. Hence, we were obligated to use the dataset containing TI-weighted MRI scans of infected patients with three kinds of brain tumor: meningioma, glioma, and pituitary tumor. Deep learning is a kind of machine learning that uses neural networks to simulate how people learn subjects [3]. Deep learning makes it quicker and simpler to collect, analyze, and interpret vast amounts of data., which is very advantageous for those who are entrusted with doing so. We are using such deep learning techniques to perform image segmentation on MRI scans of the brain [4]. Image segmentation is the process of dividing up images into several segments and grouping those that belong to the same object class together. By segmenting an image, complexity of classification can be reduced and/or the representation of the image can be changed to make it more meaningful and clearer. These techniques are to be used to segment the MRIs in such a way that the tumor is recognizable as a part of the brain. The domain of medical image segmentation has been incentivized with the engenderment of Convolutional Neural Networks and encoder-decoder type architectures like U-Net, where the focus is specifically on preserving and generalizing localization of Regions of Interest. An architecture like U-Net preserves both the spatial and localized features of the input image in a highly optimized manner. Furthermore, techniques like Attention mechanism, Residual Blocks, Backbones, Recurrent networks and more, have also proved to work efficiently for segmentation related tasks. Our study seeks to do image segmentation on brain MRI data using deep learning techniques, for faster and more accurate cancer identification and localization in the brain. Patients, specifically over the age of 40 years, with brain tumors have a dismal survival rate, and many tumors even go unnoticed. It takes extremely little time to forecast a brain tumor when these algorithms are applied to MRI pictures, and higher accuracy makes it easier to treat patients where complicated cases usually require experienced medical personnel to locate the area of the tumor, compare affected tissues with nearby regions, and provide the final verdict. The radiologists can make speedy decisions thanks to these projections.

The major contribution of the paper is mentioned as below:

1) Segmentation of the MRI images using deep learning algorithms for better generalization capabilities.
2) Investigation of enhanced deep learning segmentation algorithms for brain tumors.

The rest of the paper is organized as Section 2 discusses the related literature present in the domain and motivation for doing this work. Section 3 presents the methodology used in the work for brain tumor segmentation. Section 4 provides the details of the deep-learning hyper-parameters and the evaluation metrics used in this work. Also, this section presents the statistical and visual result along with a discussion. Section 5 concludes the paper.

## 2. Related Works

Most studies show analysis on image data or classification techniques with remote segmentation techniques. Javaria Amin et al. [5] is an overview of the latest techniques used in deep learning that are being used for tumor detection. The paper delves deeply into the types of imaging involved for the brain, such as MRIs, CAT scans, PET etc. Different image segmentation techniques need to be used for different types of imaging methods. The paper also gives us an insight about the datasets publicly available, and the criteria by which each deep learning model used for image segmentation was evaluated. This research helped us identify that we are to use MRI imaging data for our own research and that deep learning models like U-Net and its variants would be best suited to get the most accurate results. Instead of using the resulting tumor segmentation masks, Getty N. et al. [6] recreated the capsule network architecture [7] , refined the model parameters and MRI tumor images' preprocessing. It is stated that segmentation is not required for identifying the type of tumor, but the laborious problem of manually segmenting the tumor for localization and description still remains at hand. For mass detection in MRI scans, BrainMRNet, a deep learning model was developed by Toğaçar et al. [8]. Interestingly, an additional segmentation technique is also developed which determines the lobe area in the brain with higher concentration of two classes of tumors. Gunasekara SR et al. [9] present a comprehensive end-to-end systematic method for MRI-based tumor segmentation and detection of meningiomas and gliomas. A straightforward CNN algorithm is used to classify brain tumors, then a Faster R-CNN network is used to localize the tumor, and finally the Chan-Vese algorithm [10] is used to precisely segment the tumor. The final output was the precise tumor boundary for segmentation purposes for each given axial brain MRI. All three algorithms were linked in a cascade fashion. The suggested Faster R-CNN model is used to extract the bounding box, and then a segmentation approach is used to provide the tumor's precise contour. For early brain tumor detection, a brilliant approach was suggested by Suneetha and Rani [11]. The proposed method involves pre-processing the acquired brain MRI images using the Optimized Kernel Possibilistic C-means Method (OKPCM). To enhance the image, an adaptive Double Window Modified Trimmed Mean Filter (DWMTMF) is then employed. At last, the images are segmented using the region expanding method. Kadkhodai et al. [12] created an image enhancing model. Based on the intensities, the enhanced images are then segmented using 3D super-voxels. The study proposes a saliency detection-based feature employed with an edge-aware filtering technique which aligns the edges of the original image and saliency map further enhancing the border of the tumor. The output of their neural network is the segmented tumor. A proposition of Enhanced Convolutional Neural Network (ECNN) was put forth by Thaha et al. [13] accompanied by BAT algorithm as the loss function with the primary aim of presenting optimization-based segmentations. Small kernels used in the model, when network has lower weights, enable deep architectural design and have favorable effects on overfitting.

Ronneberger et al [14] proposed an architecture with a symmetric pair of contracting path to collect context and an expanding path to enable exact localization. In certain categories, the study easily won the 2015 ISBI Cell Tracking Challenge. The work of Oktay et al. [15] learns to concentrate automatically on target structures of varying forms and sizes in medical imaging by combining the attention gate model with the U-Net architecture. According to experimental findings, Attention gates preserve computational efficiency while continuously enhancing U-Net's accuracy across various datasets. Zhang et al. [16], suggest the use of ResUnet in their research. This study proposes a neural network for semantic segmentation that blends residual learning with U-Net. This model has the advantage of making deep network training easy due to residual units and the quantity of skip links within the network that enable information flow. The concept is commonly used to partition medical images into different bodily areas and modalities. [17][18]. To overcome the difficulty of distinguishing healthy cells from tumor boundaries in the diagnosis of brain tumors, DeepSeg, a new deep learning architecture has been developed by Zeineldin et al. [19]. The system that was developed is an interactive decoupling framework in which the encoder part uses a convolutional neural network to do spatial information processing and the decoder part provides the full-resolution probability map from the generated map. The dense convolutional network (DenseNet), the NASNet using modified U-Net, and the residual network (ResNet) were just a few of the CNN models included in the study. For colonoscopic image segmentation, Jha et al. [20] developed ResUNet++, an enhanced ResUNet architecture that suggests methods to boost its sensitivity to the important elements, suppress the unimportant features, and give larger context. These methods include residual blocks along with attention mechanism, Atrous Spatial Pyramidal Pooling (ASPP), and squeeze and excitation blocks. Comparing ResUNet++ to other techniques, the outcomes for the colorectal polyps were better. The suggested models make use of the Residual and Recurrent Network, and U-Net in the paper of Md Zahangir Alom et al. [21]. These structures for segmentation problems provide benefits like aids deep architecture training, superior feature representation for segmentation tasks ensured by feature accumulation and improves performance for medical image segmentation with the same amount of network parameters [22].

With the review of the related work in the field of medical image segmentation, studies that use segmentation algorithms leave potential grounds for higher performance with fewer complexities. The problem of segmenting medical images has benefited from the development of U-Net architecture by using the capabilities of attention and recurrent mechanisms, different backbones, and residual blocks. It has shown more opportunities to improve the effectiveness of such an architecture for tumor segmentation.

## 3. Materials and Methodology

As presented in Fig. 1 our study follows the ensuing methodology for acquiring results from the input dataset. The dataset is initially preprocessed to provide appropriate data for the implementations. Images obtained from preprocessing are then partitioned into training, validation, and testing datasets. After determining the appropriate hyperparameters, the designated training and validation data are used to train the model. Using the inference on validation data, the model with best training coefficients is saved. Testing data is then utilized to gather inference using this model. Comparison between different methodologies using various metrics is performed.

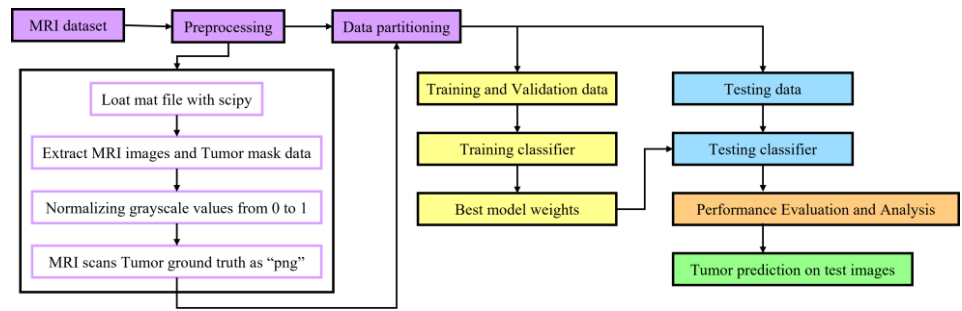

**Fig. 1.** Flow chart for conducting this work

### 3.1 Image Dataset

The brain T1-weighted CE-MRI dataset was obtained from Tianjing Medical University and Nanfang Hospital in Guangzhou, China, between 2005 and 2010 [23]. The dataset was first shared publicly in 2015 and saw multiple revisions, with the most recent iteration of the dataset released in 2017. The collection comprises 708 meningiomas, 1426 gliomas, and 930 pituitary tumors in 3064 T1-weighted, contrast-enhanced pictures, as depicted in Fig. 2, from 233 patients. Data fields for the tumor label, patient ID, image data, tumor boundary and mask data are stored in MATLAB file format. Further, a crucial preprocessing stage was required to appropriately use the data with Python.

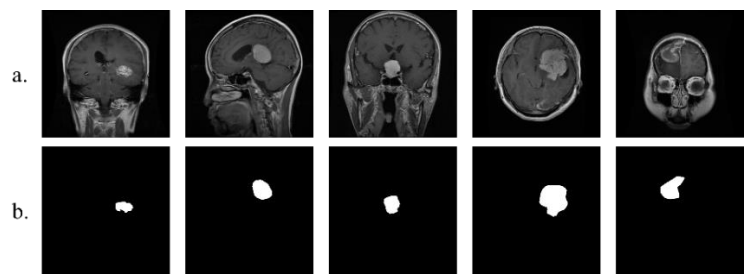

**Fig. 2.** Sample images of T1-weighted MRI dataset, a) MRI scans b) Ground Truth masks

### 3.2 Preprocessing Stage

Since the files were in MATLAB format, it was required to convert and preserve their contents as standard images so that our techniques could use them effectively without wasting memory by loading each mat file repeatedly. Using the SciPy module, the files with the "mat" extension were imported as dictionaries.

The data included fields such as tumor label : 1 for meningioma, 2 for glioma, 3 for pituitary tumor, patient id, image data: grayscale values with 3 channels in the form of an array, tumor border : a vector storing the coordinates of discrete points on tumor border; [x1, y1, x2, y2,...] in which x1, y1 indicate planar coordinates of tumor border, and tumor mask data : a binary image with 1 indicating tumor region and 0 indicating non-tumor region. Image data and tumor mask data were extracted as NumPy arrays by iterating through the mat files. The values were normalized between 0 and 1. Using the OpenCV module, all the processed images are stored in "png" format and divided into train (2485 instances), validation (274 instances), and test (305 instances) sets.

### 3.3 Methodology

**U-Net**

U-Net is an architecture for a convolutional neural network that was developed mainly for image segmentation and offers advancements over CNNs, to deal with biomedical images where the goal is to not only categorize the infection but also to identify its location [14]. Prior to U-net, classification networks were unable to segment an image using pixel-level contextual information. U-net's adaptability led researchers to heavily include it in subsequent investigations using new imaging techniques. This research has grown steadily throughout the years, incorporating several imaging techniques and application fields. [24]

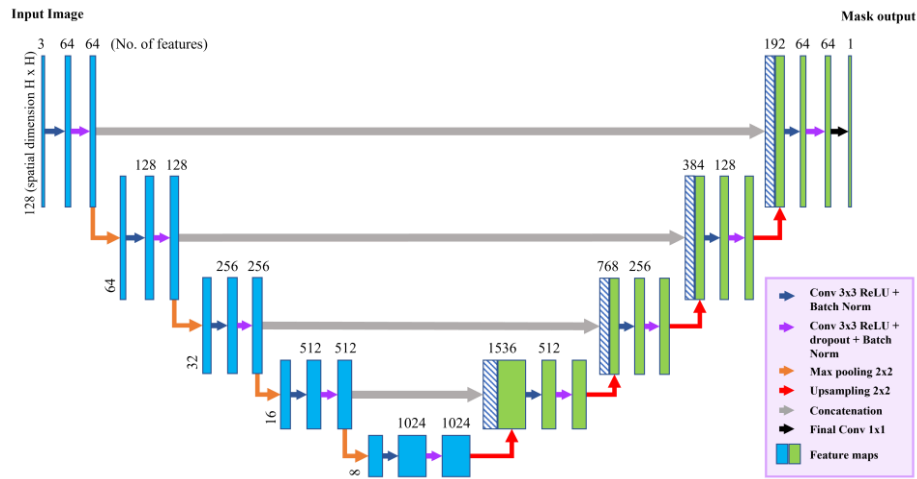

**Fig. 3.** U-net architecture for tumor segmentation

As presented in Fig. 3, the U-Net design can be seen as a pair of encoder-decoder networks with an expanding path that provides exact localization and a symmetric contracting path that collects contextual and spatial data, giving it a u-like shape. Each block of the contracting route, repeated for a few increasing filters, consists of two successive 3 x 3 convolutions, a ReLU activation unit and a max-pooling layer.

The U-Net's novel features may be found in the extending path, where, in each stage, the feature map is upsampled by 2 x 2 up-convolution. The contraction path's feature map from the matching layer is stacked onto the upsampled feature map to transport the encoder's high-resolution feature maps directly to the decoder network using skip connections. Then come two successive 3 x 3 convolutions, escorted by ReLU activation. This design not only exceeded the most effective technique at its imminence (a sliding window ConvNet) but was also simpler and quicker to train end-to-end with less input images.

### Attention U-Net

Adding to the capabilities of U-net, the attention mechanism [15] intends to recreate the capability of humans to concentrate on relevant instances while ignoring others in the neural network and allocate more computing resources to important stimuli. The task is achieved using an attention gate which when added to the skip connection within the U-Net, provides localized classification resulting in more accurate and robust image classification performance and progressively suppresses highlight responses in unrelated background areas. As shown in Fig. 4, attention gate accepts input from the succeeding deepest layer feature map (better feature representation) and skip connection feature map from the contracting path (better spatial information). A sequence of operations like, adding the inputs after convolution to obtain aligned and unaligned weights (selective concentration), activation layer, single filter and stride convolution, sigmoid function for scaling weights and finally, resampling. The output and skip connection feature map are multiplied and continue the procedure of U-Net. Gradients from background areas are de-weighted in the backward pass, this allows the model parameters in earlier layers to be modified based on spatial zones that are crucial for a particular task, and subsequently improves model sensitivity and accuracy towards foreground pixels.

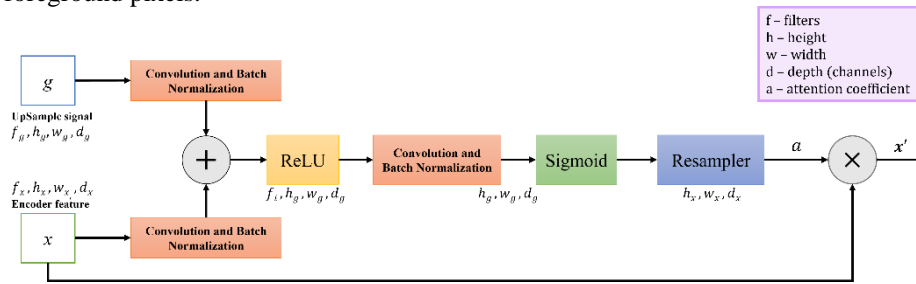

**Fig. 4.** Attention gate with gating signal (g) from encoder and current level decoder feature map (x) with X as output to be concatenated with x

### ResUnet

He et al [25]. addressed in their study that a very deep architecture is difficult to train due to issues like vanishing gradients and adverse effects on the generalization power of the model. To overcome this predicament, they suggested using an identity mapping architecture for deep residual learning. The intuition to solve the degradation problem is to recognize that shallower networks perform better than the deeper networks and skipping extra layers can help us maintain the depth of layers and prevent degradation.

In this model, the layers give an output as H(x) = F(x) + x, where F(x) is residual (difference between output and input) and x being the identity skip connection as shown in Fig. 5. Hence, the layers in the residual network are learning the residual and not the true output which resolves the vanishing gradient problem and helps the system avoid lossy compression with identity mapping.

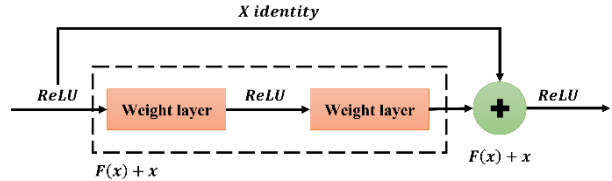

**Fig. 5.** Residual learning: a building block

Using the best of Deep Residual learning and U-Net architecture, Zhang et al. [16] came up with the Deep Residual U-Net architecture. ResUnet also consists of encoding path, skip connections and decoding path, where both the encoder and decoder units incorporate residual blocks for each level. Combining these crucial techniques, the network's training process is facilitated by the residual unit, and information spreads without degradation through the skip connections. Anita et al. [18] found in their study that, for lung segmentation, more discriminative feature representations can be extracted by a deeper residual network than a shallow network.

### ResUnet++

ResUnet++ is based on the structure of ResUnet. In addition to that, it uses techniques like squeeze and excitation blocks, attention blocks, and Atrous Spatial Pyramidal Pooling (ASPP) [20]. This model begins with a stem block which is used to downsample the input image to ensure operations maintain low computational complexity and efficiency in results and passed onto the encoding path. Each encoder block's output is passed to the squeeze-and-excitation block which modifies the features and enhances the network's characteristic power. This leads to suppression of the redundant features and increased sensitivity for pertinent features [26]. Through the ASPP network, the output of the encoding block is passed [27] which connects the encoder to the decoder. It enlarges the field-of-view of the filters by providing multi-scale information to include a broader context. Before each decoder block, an attention block is used to improve the quality of features that improves the outcome. To obtain the final segmentation map at the end of the decoding path, another ASPP block is utilized followed by a 1x1 convolution with sigmoid activation.

### Recurrent Residual U-Net (R2Unet)

Along with the structure of U-Net and the residual block technique, recurrent convolutions improve the model's capability to integrate context information to ensure better feature representation for segmentation. Keeping track of former input and current pixel information is necessary for recurrent blocks to anticipate future output. This helps in remembering and integrating context information which is most significant in semantic segmentation [28].

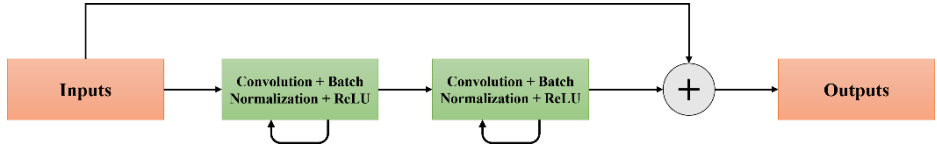

**Fig. 6.** Recurrent Residual convolutional units (RRCU)

The combination of residual units and recurrent units as depicted in Fig. 6 are extended to each block of the encoder and decoder component of U-Net architecture [21]. The residual unit helps in avoiding performance degradation by incorporating long/short skip connections over deep networks, and the recurrent unit can withhold reasonable dependencies among pixel values by considering contextual data. In blocks of both the encoder and decoder, recurrent convolutional layers (RCLs) with residual units are utilized in place of conventional forward convolutional layers, which aids in the development of a deeper and effective model. This segmentation approach demonstrates the efficiency of feature accumulation from one portion of the network to another and shows benefits for both training and testing phases.

## 4. Experimentations and Result Discussion

The experimentations were carried out on the Jupyter notebook platform. The collected data was preprocessed using SciPy to load MATLAB data and open-cv for normalizing the grayscale values and saving the images. The prepared data was trained on algorithms implemented using keras library and keras-unet-collection [29]. The experimental works, including training and testing, are carried out using the NVIDIA TESLA P100 GPU with 2 CPU cores, 16 GB GPU memory and 13 GB RAM. All models in our study are trained entirely from scratch without any prior weights. Only for U-Net and Attention U-Net, three different backbones are utilized to incorporate their salient features, viz. VGG-19, ResNet152 and DenseNet201.

### 4.1 Deep learning Hyperparameters

The Table. 1 presents the deep learning hyperparameters used in each of the tumor prediction algorithms which are kept the same for exact comparison.

**Table 1.** Hyper-parameters used for training of various deep learning models

| Hyperparameters | Values |
|---|---|
| Learning Rate | 0.001 |
| Beta 1 | 0.9 |
| Beta 2 | 0.999 |
| Optimizer | Adam |
| Loss function | Binary Cross Entropy |
| Batch Size | 32 |
| Epochs | 100 |

### 4.2 Evaluation Metrics

A trained model's results should be summarized using metrics that are better at showing the model's segmentation skills. The precision (P) is calculated using $P = \frac{tp}{tp + fp}$, recall (R) is calculated using $R = \frac{tp}{tp + fn}$ which are used to calculate F1-Score represented by $F1\ Score = 2 * \frac{P * R}{P + R}$ and IoU is given as $IoU = \frac{tp}{tp + fn + fp}$. In the formulae, $tp$ represents true positive, $tn$ represents true negative, $fp$ represents false positive and $fn$ represents false negative. These metrics are often utilized for judging the performance of medical image segmentation [30], [14], [15] and are preferred over pixel accuracy as they are better at measuring the segmentation's perceptual quality [31]. For object segmentation, the dice score more accurately measures size and localization agreement [32].

### 4.3 Statistical and visual results

Based on the evaluation metrics, both statistical and visual results were gathered using the predicted output and ground truth which are provided below.

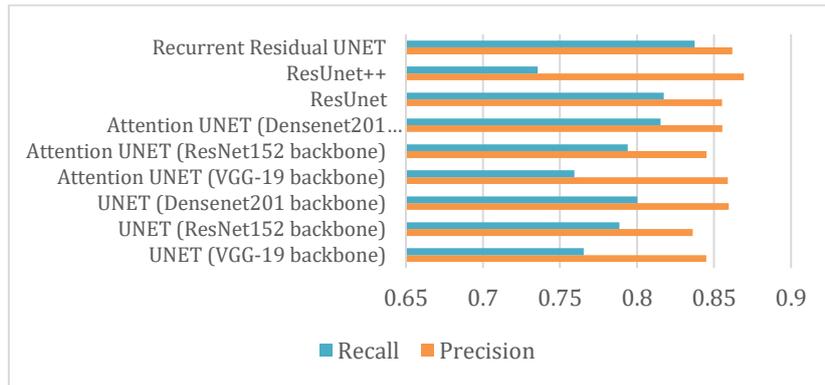

**Fig. 7.** Precision and Recall of all applied deep learning models for tumor segmentation

**Table 3.** Statistical results of various state-of-art deep learning models for tumor segmentation

| Methodologies | F1 Score | Mean IoU |
|---|---|---|
| UNET (VGG-19 backbone) | 0.8033 | 0.8322 |
| UNET (ResNet152 backbone) | 0.8116 | 0.8382 |
| UNET (Densenet201 backbone) | 0.8288 | 0.8507 |
| Attention UNET (VGG-19 backbone) | 0.8060 | 0.8342 |
| Attention UNET (ResNet152 backbone) | 0.8188 | 0.8434 |
| Attention UNET (Densenet201 backbone) | 0.8349 | 0.8553 |
| ResUnet | 0.8360 | 0.8562 |
| ResUnet++ | 0.7969 | 0.8272 |
| **Recurrent Residual UNET** | **0.8495** | **0.8665** |

| Instance Number | MRI Scan | Ground Truth | U-Net | Attention U-Net | ResUnet | ResUnet++ | R2Unet |
|---|---|---|---|---|---|---|---|
| 1. | 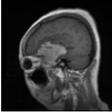 | 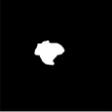 | 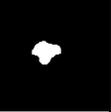 | 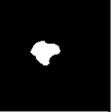 | 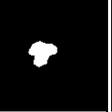 | 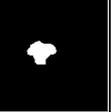 | 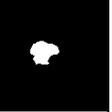 |
| 2. | 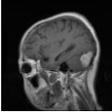 | 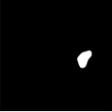 | 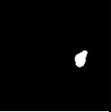 | 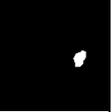 | 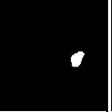 | 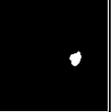 | 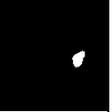 |
| 3. | 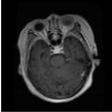 | 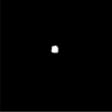 | 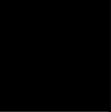 | 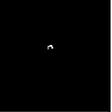 | 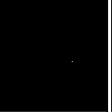 | 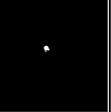 | 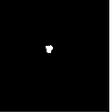 |
| 4. | 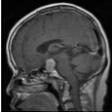 | 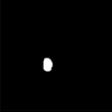 | 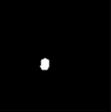 | 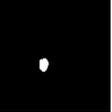 | 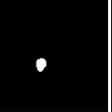 | 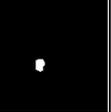 | 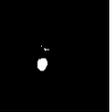 |
| 5. | 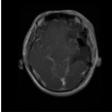 | 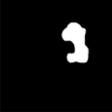 | 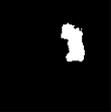 | 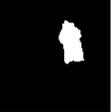 | 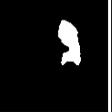 | 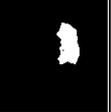 | 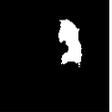 |
| 6. | 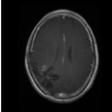 | 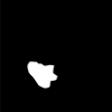 | 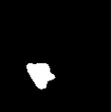 | 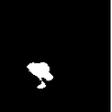 | 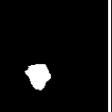 | 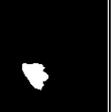 | 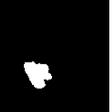 |
| 7. | 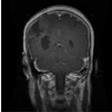 | 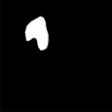 | 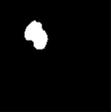 | 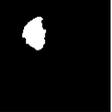 | 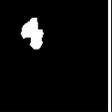 | 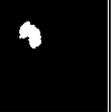 | 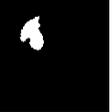 |
| 8. | 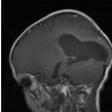 | 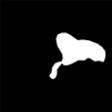 | 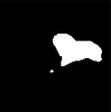 | 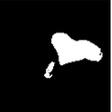 | 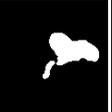 | 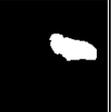 | 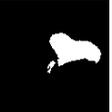 |
| 9. | 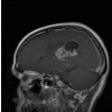 | 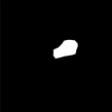 | 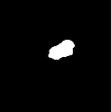 | 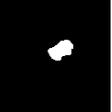 | 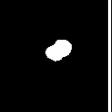 | 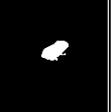 | 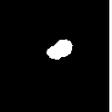 |
| 10. | 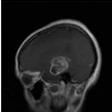 | 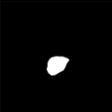 | 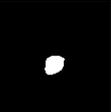 | 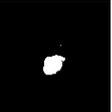 | 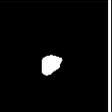 | 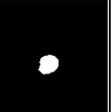 | 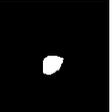 |

**Fig. 8.** Visual (Qualitative) results of various brain tumor segmentation models on T1-Weighted MRI scans

Based on the metrics, all the models quantitatively performed well. Table 3 and Fig. 7 findings show that U-Net and Attention U-Net are more accurate with DenseNet201 as the backbone. Primarily, it can be concluded from statistical findings in Table 3 that Recurrent Residual UNET (R2Unet) is the most effective model for segmenting brain tumors from the test data. R2Unet (highlighted red in Table 3) outperformed other deep learning models according to all metrics except precision where ResUnet++ seems to perform better. For comparing qualitative results of the models, we chose 10 random samples from the test dataset as shown in Fig. 8. These MRI images exist in the form of three different planes, viz. Sagittal, Axial and Coronal planes. Compared to the ground truth and the outcome of various models having certain defects like under and over-segmentation and erroneous borders, segmentation predictions from R2Unet are almost perfect. Instance 3 from Fig. 8 shows that U-Net and ResUnet failed to detect the tumor, while Attention U-Net and ResUnet++'s predictions are under-segmented. R2Unet, conversely, delivered accurate segmentation with little disorder in the boundary. In instances 2, 5, 6, 7, and 10, R2Unet predictions have more fidelity to the ground truth than other models. However, R2Unet's predictions occasionally (instances 1, 4, 9) show slight over-segmentation whereas ResUnet and ResUnet++ predictions are more helpful. Instances 2, 3, 4 and 10 have simple ground truth shapes, and yet ResUnet, ResUnet++ and R2Unet show jagged segmentations while both U-Net and Attention U-net retain some smoothness. Additionally, other instances (1, 5, 6, 7, 8 and 9) make it clear that the shape of the tumor is preserved by ResUnet, ResUnet ++ and R2Unet more than U-Net and Attention U-Net.

R2Unet has proved to provide better results among the compared models since it is capable of keeping a record of previous input and current pixel information to utilize in predicting future output; in this way, it helps the model integrate context information that is significant in semantic segmentation. Additionally, the introduction of the feature accumulation technique in recurrent convolutional layers has provided more robust feature representation vital for extracting low-level features especially pertinent to medical image segmentation.

## 5. Conclusion

Detecting tumors is difficult and expensive in the modern world since it is done mostly through imaging and via specialists. This can be tackled by the means of computer-aided detection. In this work, the suitable model for tumor extraction from MRI scans is investigated. The results obtained from this work show that an encoder-decoder based Convolutional Neural Network architecture fused with Recurrent and Residual units when trained on a dataset of brain tumor MRI scans, using Adam optimizer, produces a F1 score of 0.8495 and an IoU of 0.8665, and outperforms other compared models like U-Net and Attention U-Net, deep Residual U-Net with its variants. Also, the qualitative results show consistent results for various representations of the MRI scans. These algorithms reduce the burden on the doctors and make healthcare more accessible and inexpensive for all. Our work helped us realize that residual block removes vanishing gradient, the attention mechanism provides a focus on essential features for segmentation. The accuracy of the suggested approach can be further improved by employing larger, more varied datasets, and various preprocessing techniques. We are also exploring other architectures that would bridge the semantic gap between encoder and decoder feature maps by creating dense skip connections on skip pathways which would improve gradient flow. Combining all these efforts, we look forward to creating a highly accurate model for tumor segmentation.